\documentclass{elsart4}

\usepackage{amssymb}

\usepackage{graphics}
\usepackage{graphicx}
\usepackage{epsfig}
\usepackage{bm}
\journal{Physica B}

\usepackage{color}
\definecolor{green}{rgb}{0.0, 0.5, 0.0}

\usepackage{ulem}

\begin{document}

\begin{frontmatter}

% Title, authors and addresses

% use the thanksref command within \title, \author or \address for footnotes;
% use the corauthref command within \author for corresponding author footnotes;
% use the ead command for the email address,
% and the form \ead[url] for the home page:
%seo0802
%\title{Effects of charge order and spin frustration on the spin ordering in TMTTF$_2$ X}
\title{Spin frustration, charge ordering, and enhanced antiferromagnetism in TMTTF$_2$SbF$_6$}
\author[aist,uvt]{Kazuyoshi Yoshimi},\author[riken,CREST]{Hitoshi Seo}, \author[aist]{Shoji Ishibashi}, and \author[UCLA]{Stuart E. Brown}
\address[aist]{Nanosystem Research Institute ``RICS", AIST, Ibaraki 305-8568, Japan}
\address[uvt]{Department of Physics, University of Tokyo, Tokyo 113-8656, Japan}
\address[riken]{Condensed Matter Theory Laboratory, RIKEN, Saitama 351-0198, Japan}
\address[CREST]{JST, CREST, Saitama 351-0198, Japan}
\address[UCLA]{Department of Physics and Astronomy, UCLA, CA 90095, USA}

% \corauth[cor1]{Tel: +0 000 000 000; FAX:+0 000 000 000; e-mail: author@institution.xx}
% \address{Address\thanksref{label3}}
% \thanks[label3]{}

\begin{abstract}
We theoretically investigate the effects of charge order and spin frustration on the spin ordering in TMTTF salts.
Using first-principles band calculations, we find that a diagonal inter-chain transfer integral $t_{q1}$, which causes spin frustration between the inter-chain dimers in the dimer-Mott insulating state, strongly depends on the choice of anion. Within the numerical Lanczos exact diagonalization method, we show that the ferroelectric charge order changes the role of $t_{q1}$ from the spin frustration to the enhancement of the two-dimensionality in spin sector. The results indicate that $t_{q1}$ assists the cooperative behavior between charge order and antiferromagnetic state observed in TMTTF$_2$SbF$_6$.
\end{abstract}

\end{frontmatter}

% main text
\section{Introduction}
Low-dimensional molecular conductors provide a fruitful stage to study strong electron correlation effects leading to a wide variety of phase transitions~\cite{review}. In this context, among the most-studied families are the quasi-one dimensional (Q1D) molecular conductors TMTTF$_2$$X$ (TMTTF:  tetramethyl-tetrathiofulvalene, $X$: monovalent anion)~\cite{review}. These salts form a Q1D $\pi$-band at quarter-filling in terms of holes with intrinsic dimerization along the conduction axis. They exhibit various types of phase transitions such as ferroelectric charge ordering (FCO), spin-Peierls (SP), antiferromagnetic (AF), and superconducting (SC) transitions by applying pressure or replacement of $X$~\cite{review2,Yu04,Itoi08,Iwase09}. Among them, (TMTTF)$_2$SbF$_6$ shows a peculiar behavior under pressure; a cooperative reduction of FCO and AF phase transition temperatures by the application of pressure has been reported by NMR measurements~\cite{Yu04}. This result naively does not coincide with the case for typical CO transitions, where CO suppresses the tendency toward magnetic ordering due to decrease of the effective spin exchange couplings~\cite{seo06,Otsuka05}. Recently we have proposed a mechanism where the FCO transition increases the two-dimensionality in the magnetic sector and leads to the stabilization of the AF state~\cite{condmat}. In this study, we have also found that the diagonal inter-chain hopping $t_{q1}$ strongly depends on the choice of anion. Intriguingly, $t_{q1}$ causes geometrical frustration between the chains of spins localized on every dimer, forming Q1D exchange couplings as seen from Fig.~\ref{Fig.1} (b); we will see this effect more explicitly later.

In this paper, we investigate the role of $t_{q1}$ for the spin ordering in the absence or the presence of FCO, and how it affects the cooperative behavior between FCO and two-dimensional (2D) AF states in TMTTF$_2$SbF$_6$.

\section{Formulation}
We investigate a Q1D extended Hubbard model at 1/4-filling in terms of holes with Coulomb interactions and inter-chain hoppings, whose Hamiltonian is given by
 \begin{eqnarray}
{ \mathcal H }_{\rm EHM}=-\!\sum_{\langle i j\rangle, \sigma} t_{ij}(c_{i\sigma}^{\dag}c_{j\sigma}+{\rm H.c.})\nonumber\\
\hspace{20mm} +U\!\sum_{i}n_{i\uparrow}n_{i\downarrow} +\sum_{\langle ij\rangle}V_{ij}n_{i}n_{j},
\label{EHM}
\end{eqnarray}
where $t_{ij}$ is the transfer integral between the neighboring sites denoted by $\langle i j \rangle$,
$c_{i \sigma}^{\dag}$ ($c_{i \sigma}$) is the creation (annihilation) operator of a hole on the $i$th site with spin $\sigma=\uparrow$ or $\downarrow$, and  $n_{i}=n_{i\uparrow}+n_{i\downarrow}$ with $n_{i\sigma}=c_{i\sigma}^{\dag}c_{i\sigma}$. $U$ and $V_{ij}$ are the on-site and the inter-site Coulomb interactions, respectively.

\begin{figure}
\begin{center}
\includegraphics[width=6cm]{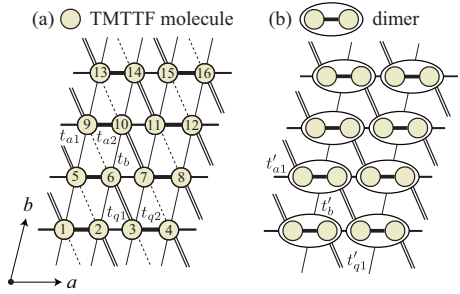}
\end{center}
\vspace*{-1em}
\caption{ The schematic representation of the structures in the donor plane for (a) TMTTF compounds and (b) dimers composed of two TMTTF molecules. In dimerized pictures, the transfer integrals are effectively given by $t_{a1}'=t_{a1}/2,~t_{b}'=t_b-t_{q2}/2,$ and $t_{q1}'=t_{q1}/2$, respectively.}
\label{Fig.1}
\end{figure}

The transfer integrals have been estimated~\cite{condmat} for two representative members, (TMTTF)$_2$PF$_6$ and (TMTTF)$_2$SbF$_6$ (hereafter, we call them as PF$_6$ salt and SbF$_6$ salt) by the first-principles band calculations using the computational code QMAS (Quantum MAterials Simulator)~\cite{QMAS} based on the projector augmented-wave method~\cite{PAW} with the generalized gradient approximation~\cite{GGA}. We determine the values by fitting to the band structures.

The values of transfer integrals are given as $\{t_{a1}, t_{a2}, t_{b}, t_{q1}, t_{q2} \}$
$=$$\{-155, -203, 26.2, -1.31, -3.29\}$ for PF$_6$ salt and $\{-149, -207, 16.4, -16.4, -9.73\}$ for SbF$_6$ salt, in the unit of meV (notations are shown in Fig.~\ref{Fig.1} (a) ). From these results, we can see that the diagonal hoppings $t_{q1},t_{q2}$ are negligible for PF$_6$ salt but appreciably large for SbF$_6$ salt, especially $|t_{q1}/t_b|\sim 1$.

Considering these results, we hereafter set the transfer integrals as $t_{a1}=-0.8,~t_{a2}=-1,~t_{b}=0.15,~t_{q2}=0$ and choose the diagonal inter-chain transfer integral $t_{q1}$ as a parameter. We set the on-site Coulomb interaction as $U=4$ and impose a constraint on $V_{ij}$ as $V_{a1}=V_{a2}=V_{q1}=V_{q2}=V$ and $V_b=0$ to realize the FCO pattern shown in Fig.~\ref{Fig.3} (b).

We perform numerical exact diagonalization on a $4\times4$ sites cluster under the periodic boundary condition and calculate the following inter-dimer/intra-dimer charge and spin structure factors given by
\begin{eqnarray}
\hspace{10mm}C_{\pm}(\bm q)&=& \frac{1}{N_d}\sum_{i, j}\langle n_{i}^{\pm}n_{j}^{\pm}\rangle {\rm e}^{i {\bm q}\cdot({\bm r}_i-{\bm r}_j)}, \\
S_{\pm}(\bm q)&=& \frac{1}{N_d}\sum_{i, j}\langle m_{i}^{\pm}m_{j}^{\pm}\rangle {\rm e}^{i {\bm q}\cdot({\bm r}_i-{\bm r}_j)},
\end{eqnarray}
where $N_d$ is the total number of dimers and ${\bm r}_i$ denotes the center position of the $i$th dimer.
Here, the inter-dimer/intra-dimer ($+$/$-$) correlations are detected by the summation/difference in charge and spin densities within each dimer, $n_{i}^{\pm} =(n_{2i}\pm n_{2i+1})/2$ and $m_{i}^{\pm} =(m_{2i}\pm m_{2i+1})/2$ with $m_{i}=n_{i\uparrow}-n_{i\downarrow}$, respectively, where even (odd) number is labeled as the site for the left (right) side in a dimer, as shown in Fig.~\ref{Fig.1} (a).

\section{Results}
\begin{figure}
\begin{center}
\includegraphics[width=7.5cm]{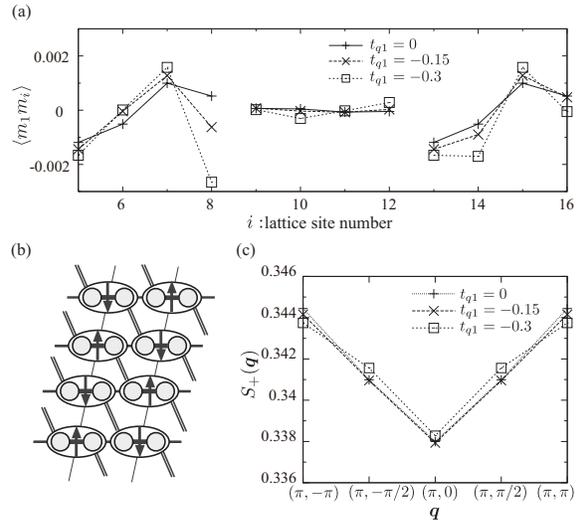}
\end{center}
\vspace*{-1em}
\caption{ (a) Inter-chain real space spin-spin correlation functions between $1$ and $i$ sites $\langle m_1 m_i \rangle$ (the site numbers are shown in Fig.~\ref{Fig.1} (a)) at $V=0$ for $t_{q1}=0,~-0.15$, and $-0.3$. Lines are connected between sites in a same chain.
(b) A schematic figure for 2D AF state between dimers. (c) Inter-dimer spin structure factors $S_+(\pi, q_y)$ at $V=0$ for $t_{q1}=0,~-0.15$, and $-0.3$. Lines are guides for eyes.}
\label{Fig.2}
\end{figure}
First, we show the inter-chain real space spin-spin correlations $\langle m_1 m_i\rangle$ at $V=0$ in Fig.~\ref{Fig.2}(a). The system is in the dimer-Mott insulating regime~\cite{condmat}, where the intra-chain dimerization and $U$ lead to a charge-uniform Mott insulator. For $t_{q1}=0$, the sign of $\langle m_1 m_i\rangle$ changes each other between dimers; there exists AF correlation as shown in Fig.~\ref{Fig.2}(b), although the inter-chain correlation is weak due to small $t_b$. 
With increasing $|t_{q1}|$, this AF correlation is diminished, as seen, for example, in the sign change of $\langle m_1 m_8\rangle$, for which the sites are connected by $t_{q1}$. As a result, the AF correlation between inter-chain dimers is weakened. In Fig.~\ref{Fig.2}(c), we show the inter-dimer spin structure factors $S_+(\pi, q_y)$. It has peaks at $q_y= \pm \pi$, indicating AF correlation between dimers. With increasing $|t_{q1}|$, $S_+(\pi, \pm \pi)$ are suppressed, while $S_+(\pi, \pm \pi/2)$ and $S_+(\pi, 0)$ are enhanced; $t_{q1}$ operates as the inter-chain spin frustration parameter.

\begin{figure}
\begin{center}
\includegraphics[width=7.5cm]{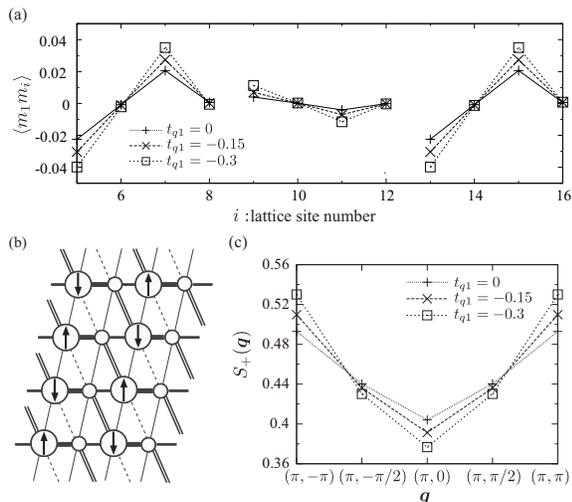}
\end{center}
\vspace*{-1em}
\caption{ (a) Inter-chain real space spin-spin correlation functions between $1$ and $i$ sites $\langle m_1 m_i \rangle$ (the site numbers are shown in Fig.~\ref{Fig.1} (a)) at $V=2$ for $t_{q1}=0,~-0.15$, and $-0.3$. Lines are connected between sites in a same chain. (b) A schematic figure for 2D AF state in FCO state. The size of a circle represents the hole density. (c) Inter-dimer spin structure factors $S_+(\pi, q_y)$ at $V=2$ for $t_{q1}=0,~-0.15$, and $-0.3$. Lines are guides for eyes.}
\label{Fig.3}
\end{figure}
Next, in Fig.~\ref{Fig.3}(a), we show $\langle m_1 m_i\rangle$ at $V=2$ for the CO region. For $t_{q1}=0$,  $\langle m_1 m_i\rangle$ is large (note the difference in the scale from Fig.~\ref{Fig.2}(a)) only at odd numbers $i_{\rm odd}$ and its sign changes each other between the neighboring charge rich sites, indicating the enhanced development of 2D AF correlations in the presence of FCO, as shown in Fig.~\ref{Fig.3}(b)~\cite{condmat}. When $t_{q1}$ is introduced, the absolute value of  $\langle m_1 m_{i_{\rm odd}}\rangle$ becomes appreciably larger, in contrast to the case of $V=0$. Such a behavior is also manifested in $S_{+}(\pi, q_y)$, as shown in Fig.~\ref{Fig.3} (c), where $S_+ (\pi, \pm\pi)$ is seen to become large with increasing $|t_{q1}|$, while $S_+(\pi, \pm \pi/2)$ and $S_+(\pi, 0)$ become small. These results indicate that $t_{q1}$ enhances the two-dimensionality in magnetic sector and further stabilizes the 2D AF state.

\begin{figure}
\begin{center}
\includegraphics[width=5.5cm]{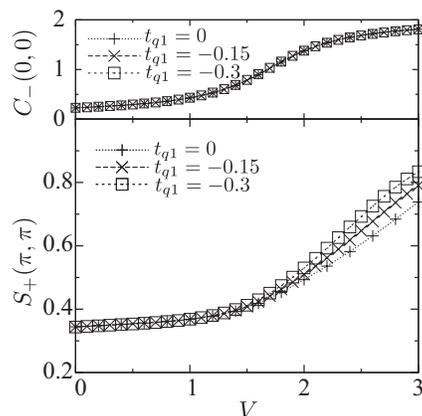}
\end{center}
\vspace*{-1em}
\caption{ (a) Intra-dimer charge structure factors $C_-(0, 0)$ and (b) inter-dimer spin structure factors $S_+(\pi,\pi)$ as functions of $V$ for $t_{q1}=0,~-0.15$, and $-0.3$. Lines are guides for eyes.}
\label{Fig.4}
\end{figure}

The role of $t_{q1}$ on the 2D AF correlation depends on the degree of CO correlation. In Fig.~\ref{Fig.4}, $C_{-}(0, 0)$ and $S_{+} (\pi, \pi)$ are plotted as a function of $V$ for $t_{q1}=0,~-0.15$, and $-0.3$. For each $t_{q1}$, we see that $C_-(0, 0)$ and $S_+(\pi, \pi)$ cooperatively develops in $V>1.5$; the FCO correlation assists in stabilizing the AF state. The noticeable point is that $t_{q1}$ becomes relevant for the AF correlation in the FCO state, as seen from the enhancement of $S_+ (\pi, \pi)$ by $t_{q1}$ only for $V>1.5$, whereas the change is small for $V<1.5$. On the other hand, $t_{q1}$ is irrelevant for the FCO correlation, i.e. $C_-(0, 0)$ does not depend on $t_{q1}$.

\section{Discussion and Summary}
For insight into the effects of $t_{q1}$ on the magnetic properties, we consider the strong coupling limits and estimate the leading terms of the spin exchange couplings by perturbation calculations with respect to the transfer integrals. In the dimer-Mott insulating state for the case of large $U$ and small $V$, the spin-exchange couplings between dimers along $a$-axis and $b$-axis are given by $J_a=-t_{a1}^2/U_d$ and $J_b=-4t_{b}^2/U_d$ with the effective on-site Coulomb interaction for the dimer units $U_d$~\cite{seo06}. Also, in the presence of $t_{q1}$ there is a diagonal inter-chain spin-exchange coupling $J_{ab}=-t_{q1}^2/U_d$. This term causes the spin frustration between inter-chain dimers as seen from Fig.~\ref{Fig.1} (b) and is enlarged toward $2t_b \sim |t_{q1}|$; in fact, the $q_y$ dependence of $S_+(\pi, q_{y})$ becomes weak with increasing $|t_{q1}|$ as shown in Fig.~\ref{Fig.2} (c).

On the other hand, in the basis of the FCO state for the case of large $U$ and $V$, the coupling between the nearest-neighbor intra-chain charge rich sites is given by $J_a \sim -4 t_{a1}^2 t_{a2}^2 / (9 UV^2)$ from the forth-order perturbation~\cite{Ohta94}. For $t_{q1}=0$, the coupling between inter-chain charge rich sites is given by $J_b\sim -4t_b^2/U$ from the second-order perturbation. Since $J_a$ is suppressed with increasing $V$, $J_b$ can become the same order compared to $J_a$. This gives the increase of the two-dimensionality in the magnetic sector and induces 2D AF state~\cite{condmat}. In the presence of $t_{q1}$, there is another additional contribution for the coupling between inter-chain charge rich sites $J_{b}'$ from the third-order perturbation process (for example, $1\to8\to5\to1$ and $1\to5\to8\to1$ in Fig.~\ref{Fig.1} (a)), which is given by $J_b' \sim -8t_{q1}t_{a1}t_b/(3UV)$. Here, the minus sign comes from the minus term coupling with the transfer integrals in Eq.~(\ref{EHM}). Since $t_{a1}$ is large compared to $t_{b}$, this process can become the same order compared to $J_b$ in the intermediate $V$ region. As a result, this contribution becomes relevant in the presence of FCO and the two-dimensionality in the spin sector increases. This agrees with the behavior of the $V$ dependence of $S_+(\pi, \pi)$ shown in Fig.~\ref{Fig.4} (b). We can conclude that not only the enhancement of the two-dimensionality in spin sector by FCO but also the largeness of $t_{q1}$ in SbF$_6$ salt compared to other salts promotes stabilization of the coexisting AF and FCO state observed in the experiments~\cite{Yu04}.

Finally, we comment on the role of $t_{q1}$ in the presence of electron-phonon couplings, which is relevant for the SP state in both the dimer-Mott and the CO states~\cite{Otsuka05,condmat,Clay03,Kuwabara03}. Previous works show that, in general, the SP state is suppressed and the 2D AF state is stabilized with increasing the inter-chain spin exchange coupling $J_b$~\cite{Inagaki83}. 
In PF$_6$ salt, 
it is experimentally indicated that, 
at finite-temperatures above the first order boundary between the two states, 
a quantum critical behavior is seen, controlled by a hidden quantum critical point~\cite{Chow98}. 
The competition between the SP and the 2D AF states occurs as well in the presence of FCO~\cite{Yu04,condmat}.
The role of $t_{q1}$ 
may enhance the strong quantum fluctuations and bring about difficulties in determining the magnetic phases in the FCO state at low temperatures. 
One possibility is that the fluctuation leads to the destabilization of ordered phases in the critical region, 
and results in a spin-liquid state down to the ground state, in between the SP and the 2D AF quantum critical points. 
This $t_{q1}$ effect on the magnetic properties with electron-phonon couplings remains as a future problem.

In summary, we have investigated the effects of charge order and the diagonal inter-chain transfer integral on the spin ordering in TMTTF$_2X$. The diagonal $t_{q1}$ works as spin frustration between the inter-chain dimers in the dimer-Mott insulating state and reduces the AF correlation. On the other hand, in the presence of FCO, this spin frustration is released and 2D AF correlation is developed; $t_{q1}$ enhances the two-dimensionality in the spin sector. This assistance of $t_{q1}$ for the cooperative behavior between the FCO and the AF states agrees with the experiments in TMTTF$_2$SbF$_6$.

\section*{Acknowledgments}
We thank M. Ogata and Y. Otsuka for discussions and
K. Furukawa and T. Nakamura for providing us the crystal structure data.
This work was supported by a Grant-in-Aid for Scientific Research
in the Priority Area of Molecular Conductors (No. 20110003, 20110004)
from MEXT, Japan. This material is based upon work supported in part by
the National Science Foundation under Grant Nos. DMR-0804625 and DMR-1105531,
and a Grant-in-Aid for Scientific Research (No. 20110003, 20110004, 21740270)
from MEXT, Japan.

\end{document}